\documentstyle[12pt,aasms4]{article}
\def\etal   {{et~al.}\ }

\def\zsun{{\rm\,Z_\odot}}
\def\msun{{\rm\,M_\odot}}

\def\vol#1  {{{#1}{\rm,}\ }}
\def\lya{{\rm Ly}\alpha}

\def\etal{et al.\ }

\newcount\refno
\newcount\rfno
\def\eq{$^{\the\refno\ }$\advance\refno by 1}
\def\ad{\advance\rfno by 1}

\def\clock{\count0=\time \divide\count0 by 60
     \count1=\count0 \multiply\count1 by -60 \advance\count1 by \time
     \number\count0:\ifnum\count1<10{0\number\count1}\else\number\count1\fi}

\begin{document}
\title{Cosmic Chemical Evolution}
\author{Renyue Cen\altaffilmark{1} and Jeremiah P. Ostriker\altaffilmark{2}}
\altaffiltext{1} {Princeton University Observatory, Princeton University, Princeton, NJ 08544; cen@astro.princeton.edu}
\altaffiltext{2} {Princeton University Observatory, Princeton University, Princeton, NJ 08544; jpo@astro.princeton.edu}

\begin{abstract}
Numerical simulations of standard cosmological scenarios have
now reached the degree of sophistication required to provide
tentative answers to the fundamental question:
Where and when were the heavy elements formed?
Averaging globally, these simulations
give a metallicity that increases 
from $1\%$ of the solar value at $z=3$
to $20\%$ at present.
This conclusion is, in fact,
misleading, as it masks the very strong dependency
of metallicity on local density.
At every epoch higher density regions have much higher
metallicity than lower density regions.
Moreover, 
the highest density regions quickly approach near solar metallicity
and then saturate,
while more typical regions slowly catch up.
These results are much more consistent with observational
data than the simpler picture (adopted by many)
of gradual, quasi-uniform increase of metallicity with time.
\end{abstract}


\section{Introduction}

One of the greatest successes of the Big Bang theory
is that its prediction 
that the primordial baryonic matter
is almost entirely composed of hydrogen and helium
with a trace amount of a few other light elements
is in detailed agreement with current observations 
(e.g., Schramm \& Turner 1998).
The heavier elements, 
collectively called ``metals",
are thought to be made at much later times
through nucleosynthesis in stars.
Metals are ubiquitous in the universe
in virtually all environments that have been observed,
including regions outside of galaxies, the intergalactic medium (``IGM"),
ranging from the metal rich 
intracluster medium to low metallicity Lyman alpha clouds.
However, metallicity (the ratio of the amount of mass
in metals to the total baryonic mass for a given region,
$M_{metals}/M_{baryons}$, divided by $0.02$ for
the Sun, $\zsun$)
is observed to be very variable.
For example, metallicity 
reaches as high as ten
(in units of the solar value, where value unity
corresponds to $M_{metals}/M_{baryons}=2\%$)
in central regions of active galactic nuclei
(Mushotsky, Done, \& Pounds 1993;
Hamann 1997; Tripp, Lu, \& Savage 1997)
but is as low as $10^{-3}$ for some halo stars in our own
galaxy (Beers 1999).
Disparity in metallicity values is also seen at high redshift.
For instance, metallicity in damped $\lya$ systems is
as high as $0.5$ and as low as $0.01$ at redshift $z\sim $
(Prochaska \& Wolfe 1997),
whereas it is about 0.01 in moderate column density $\lya$ clouds
at $z\sim 3$ (Tytler \etal 1995; Songaila \& Cowie 1996).
Low column density
Lyman alpha clouds at $z\sim 2-3$
appear to have still lower metallicity (Lu \etal 1998;
Tytler \& Fan 1994).

The question that naturally rises then is:
When were the metals made and why are they distributed as observed?
Can we understand the strong dependence of $Z/\zsun$ on the
gas density (at redshift zero)
and the comparable dependence of $Z/\zsun$ on redshift for regions
of a given overdensity?
While these are well-posed questions,
addressing them directly
is a formidable computational problem
and requires both a large dynamic range,
to ensure a fair piece of the universe to be modeled,
and sufficiently realistic physics being modeled
including gasdynamics, galaxy formation,
galactic winds and metal enrichment.
After years of continuous
improvement of both numerical techniques
and physical modeling, coupled with rapid
increase in computer power, 
we have now reached the point where this important
question can at last 
be addressed in a semi-quantitative fashion
using numerical simulations.

\section{Model}

The results reported on here are based on a new computation of 
the evolution of the gas in 
a cold dark matter model with a cosmological constant;
the model
is normalized to both the microwave background temperature
fluctuations measured by COBE (Smoot \etal 1992)
on large scales (Bunn \& White 1997)
and the observed abundance of clusters of galaxies in
the local universe (Cen 1998), 
and
it is close to both the concordance model of Ostriker \& Steinhardt (1995) 
and the model indicated by the recent
high redshift supernova results (Reiss \etal 1998).
The relevant model parameters
are: $\Omega_0=0.37$, $\Omega_b=0.049$, $\Lambda_0=0.63$, $\sigma_8=0.80$,
$H_0=70$km/s/Mpc,
$n=0.95$ and $25\%$ tensor mode contribution
to the CMB fluctuations on large scales.
Two simulations with box sizes of $L_{box}=(100,50)h^{-1}$Mpc
are made, each having $512^3$ cells and $256^3$ dark matter particles
with the mean baryonic mass in a cell being 
$(1.0\times 10^8,1.3\times 10^7)h^{-1}\msun$ and
the dark matter particle mass being
$(5.3\times 10^9,6.6\times 10^8)h^{-1}\msun$,
respectively, in the two simulations.
Output was rescaled to $\Omega_b=0.037$
to match latest observations (Burles \& Tytler 1998).
The results shown are mostly  based on the large box, while
the small box is used to check resolution effects.

The description of the
numerical methods of the cosmological hydrodynamic code 
and input physical ingredients can be found elsewhere
(Cen \& Ostriker 1999a,b).
To briefly recapitulate,
we follow three components separately and simultaneously:
dark matter, gas and galaxies,
where the last component is created continuously from the former two
during the simulations in places where real galaxies are thought
to form,
as dictated mostly by local physical conditions.
Self-consistently, 
feedback into IGM from young stars in the ``galaxies"
is allowed, in three related forms:
supernova thermal energy output,
UV photon output and mass ejection from the supernova explosions. 
The model reproduces the observed UV background as a function of redshift
and the redshift distribution of star formation
(``Madau Plot"; Nagamine, Cen \& Ostriker 1999),
among other diagnostics.
Metals are followed as a separate variable 
(analogous to the total gas density)
with the same hydrocode.
We did not fit
to the observed distributions
and evolution of metals,
but assumed
a specific efficiency of metal formation,
Subsequently rescaling the computed results
to an adopted ``yield" (Arnett 1996), 
the percentage of stellar mass 
that is ejected back into IGM as metals,
of $0.02$ (from an input value 0.06).

A word about the resolution of the simulation is appropriate here.
The conclusions drawn in 
this paper are not significantly affected by the finite resolution,
as verified by comparing the two simulations.
Let us give an argument for why this is so.
Although our numerical resolution is not sufficient to 
resolve any internal structure of galaxies,
the resolution effect should affect different regions 
with different large-scale overdensities more or less uniformly 
since our spatial resolution is uniform and our mass resolution
is good even for dwarf galaxies.
In other words, galaxy formation in our simulations
is not significantly biased against low density regions.
Thus, the distribution of the identified galaxies 
as a function of large-scale overdensity in the simulation
would be similar, if we had a much better resolution.
It is the distribution of the galaxies that determines
the distribution of metals, which is the subject of this paper.
Needless to say, we can not make detailed  modeling of
the ejection of metals from galaxies into the IGM and
this ignorance is hidden in the adopted ``yield" coefficient.
However, once the metals get out of galaxies,
their dynamics is followed accurately.
Changing the adopted yield by some
factor would change all quoted
metallicities
by the same factor 
but not alter any statements about spatial
and temporal distribution of metals.

\section{Results}

Figure 1 shows the evolution of metallicity 
averaged over the entire universe (dot-dashed curve)
and four regions with four different overdensities,
$\delta_\rho=(10^3,10^2,20,0)$, smoothed by a Gaussian window of comoving
size $0.3h^{-1}$Mpc, respectively,
that approximately correspond to clusters of galaxies,
Lyman limit and damped $\lya$ systems,
moderate column density $\lya$ clouds
and very low column density $\lya$ clouds,
at $z=(3,1.0,0.5,0)$.
The overdensity of each class of objects is defined using a Gaussian
smoothing window of radius $0.3^{-1}$Mpc, which corresponds
to a mean mass of $6\times 10^{10}h^{-1}M_\odot$. If we assume
that the DLAs are progenitors of the present day large disk galaxies,
their mass may be in the range $5\times 10^{12}-1\times 10^{13}h^{-1}M_\odot$.
Therefore a choice of overdensity of $100$ seems appropriate.
For the moderate column density Lyman alpha clouds, the choice
is somewhat less certain but small variations do not drastically
change the results.
For the very low column density Lyman alpha clouds,
the choice of the mean density should be adequate since 
the density fluctuations for these objects are small thus 
their density should be close to the mean.
For the clusters of galaxies we can use overdensity of $10^3$
or $3\times 10^3$ and it makes no difference to the results.
Note that a given class of objects is chosen to have 
a fixed comoving overdensity, not to have a fixed physical density.
This choice is made because the decrease of a factor $50-100$
of the observed meta-galactic radiation field
from $z\sim 3$ to $z\sim 0$ (Haardt \& Madau 1996),
and the increase of the comoving size of structure with time
at a fixed comoving density as $\sim (1+z)^{-1/2}$ (Cen \& Simcoe 1997)
approximately compensate for the decrease of physical 
density so a fixed comoving density approximately corresponds
to a fixed column density at different redshifts.
This applies for the last three classes of objects.
For the first class of objects (clusters of galaxies) either
choice gives comparable results, due to the fact
that metallicity saturates at the highest density (see below).

Several trends are clear.
First, metallicity is a strong function of overdensity in
the expected sense:
high density regions have higher metallicity.
Second, and more surprisingly,
the evolution of metallicity itself is a strong function
of overdensity: high density regions evolve slowly with
redshift, whereas the metallicity
in low density regions decreases rapidly towards high redshift.
Finally, the overall metallicity evolution averaged globally
differs from that of any of the constituent components.
Therefore {\it any given set of cosmic objects 
(including stars or $\lya$ forest)
cannot
be representative of the universal metallicity at all times},
although at a given epoch one may be able to identify
a set of objects that has metallicity close to the universal mean.
For example, at $z=3$, 
regions with overdensity $20$
(which roughly correspond to Lyman alpha clouds of column density
of $\sim 10^{14.0-15.0}~$cm$^{-2}$)
have metallicities very close to the global mean,
while at $z=0$, 
regions with overdensity of one hundred
(which roughly correspond to Lyman limit and damped
Lyman alpha systems)
has metallicity very close to the global mean.

It has been the conventional wisdom to expect
that, as all ``metals" are produced (but not, on balance, destroyed)
by stars, 
the metal abundance should increase with time or decrease
with increasing redshift.
What we see from Figure 1 is that there is another trend which is
as strong as or stronger than this.
Since stars are {\it relatively} overabundant in the 
densest regions, metallicity is a strongly increasing
function of density at any epoch.
This trend is observed within galaxies (with central parts being
most rich) but it is also true when one averages
over larger regions.
The gas in dense clusters of galaxies is far more metal rich than the
general IGM at $z=0$.
This trend is shown in another way in Figure 2, where
metallicity is plotted as a function of overdensity 
at four redshifts.

Let us now examine the individual components more closely
in Figure 3 with
panels (a,b,c,d) showing the metallicity
distributions for regions of overdensity 
$(10^3,10^2,20,0)$, respectively,
at four redshifts $z=(3,1,0.5,0)$.
We examine each panel in turn.
In panel (a)
we see that there is almost no evolution from redshift
zero (thick solid curve) to redshift one (dotted curve)
for metallicity of intracluster gas. 
The narrowness of the computed distributions fits observations
very well for clusters locally and at low redshift.
But we predict that the metallicity of clusters at redshift $z=3$
will be somewhat lower than their low redshift counterparts
by a factor of about three,
with the characteristic
metallicity declining to $Z\sim 0.1\zsun$.

Second, examining panel (b) for regions with overdensity
$10^2$, which roughly correspond to Lyman limit and 
damped Lyman alpha systems,
it is seen that the median metallicity 
increases only slightly from 
$z=3$ to $z=0.5$,
but there is a large range of metallicity expected of approximately $30$
at any redshift,
in very good agreement with observations over the entire
redshift range considered.

Next, panel (c) shows the integral 
distributions for regions with overdensity $20$,
that correspond to moderate
column density Lyman alpha clouds with column density
$10^{14}-10^{15}$cm$^{-2}$.
We see that the median metallicity 
increases by a factor of about $10$ from redshift
$z=3$ to $z=0$,
but with a broad tail towards the low metallicity
end at all redshift,
again in good agreement with observations.
Dav\'e \etal (1998) concluded that the metallicity
for regions of overdensity of $\sim 10$ at $z\sim 3$ 
is $10^{-2.5}$ from analysis of CIV absorption lines,
consistent with our results here.

Finally, panel (d) shows regions with overdensity $0$ (i.e, 
at the mean density) corresponding to 
the very low column density Lyman alpha clouds.
The observations are upper bounds.
But it appears that the bulk of the regions with such low density
indeed has quite low metallicity, consistent with observations.
Dav\'e \etal (1998) derived an upper bound on metallicity
for near mean density region at $z\sim 3$ 
of $10^{-3}$ from analysis of OVI absorption lines,
in agreement with our results.

\section{Conclusion}

In the simulation examined in this paper
high density regions reach
an approximately solar density first,
with lower density regions approaching that level at later epochs,
and at all epochs the variations of $Z$
with density is comparable to or larger
than the variations at a given overdensity.
This saturation of metallicity has a natural physical explanation.
Regions where the peaks of long and short waves fortuitously
coincide have the highest
initial overdensity and
the earliest significant star formation;
but, when the long waves break,
high temperature shocks form (as in the centers
of clusters of galaxies),
so that further new galaxy formation and infall onto older
systems ceases,
star formation declines (Blanton \etal 1999),
and the metallicity stops increasing.
Observationally, we know that,
in the highest density,
highest metallicity and highest temperature regions of the rich
clusters of galaxies, new star formation
has essentially ceased by redshift zero.
As a side note, that fact that metallicity depends
as strongly on density as on time implies that stellar 
metallicity
need not necessarily (anti-)correlate with the stellar
age. For example, young stars may be relatively metal poor,
as supported by recent observations (Preston, Beers \& Shectman 1994;
Preston 1994),
simply because these young stars may have formed
out of relatively lower density regions where metallicity
was low.

The picture presented here is,
in principle,
quite testable.
For example,
Steidel and co-workers  (Steidel 1993; Steidel \etal 1994)
and Savage \etal (1994) and others have found
that metal line systems observed along
a line of sight to a distant quasar
are invariably associated with galaxies
near the line of sight
at the redshift of the 
metal line system.
One would expect,
on the basis of the work presented here,
that there would be a strong statistical
trend associating higher metallicity systems
to closer galaxies,
since for these the typical overdensity is larger.
Figure 4 shows surface density contours on a slice
of $50\times 50\times 10h{-3}$Mpc$^3$
for galaxies (filled red; 
at a surface density of $31$ times 
the mean surface density of galaxies),
metals (green; at a metallicity of $0.16\zsun$)
and warm/hot gas (Cen \& Ostriker 1999a)
with $T=10^5-10^7$K (blue;
at a surface density of $6.8$ times 
the mean surface density of warm/hot gas).
Each respective contour contains 90\% of 
the total mass of the respective component.
We see that most of the green contours
contain red spots, each within a region of size approximately 
$1h^{-1}$Mpc; i.e., one would expect to 
see a normal galaxy associated with a metal line
system within a projected distance of $\sim 1h^{-1}$Mpc.
It is also seen from Figure 4 
that metal rich gas is generally
embedded in the warm/hot gas.
This may manifest itself
observationally as spectral features
that seem to arise from multiple phase gas at a similar redshift
along the line of sight.
Recent observations appear to have already
hinted this; Lopez \etal (1998),
using combined observations of
quasar absorption spectra from Hubble Space Telescope
and other ground-based telescopes,
noted that some C IV clouds are
surrounded by large highly ionized low-density clouds.
Finally, it may be pointed out that most of the metals
are in over-dense regions and these regions are generally 
relatively hot: $>10^5$Kelvin.
Therefore, they should be observable in the 
EUV and soft X-ray emitting warm/hot gas (Cen \& Ostriker 1999a).

\acknowledgments
The work is supported in part
by grants NAG5-2759 and AST93-18185, ASC97-40300.
We thank Ed Jenkins, Rich Mushotzky,
Jim Peebles, David Spergel, Michael Strauss
and Todd Tripp for discussions.

\clearpage
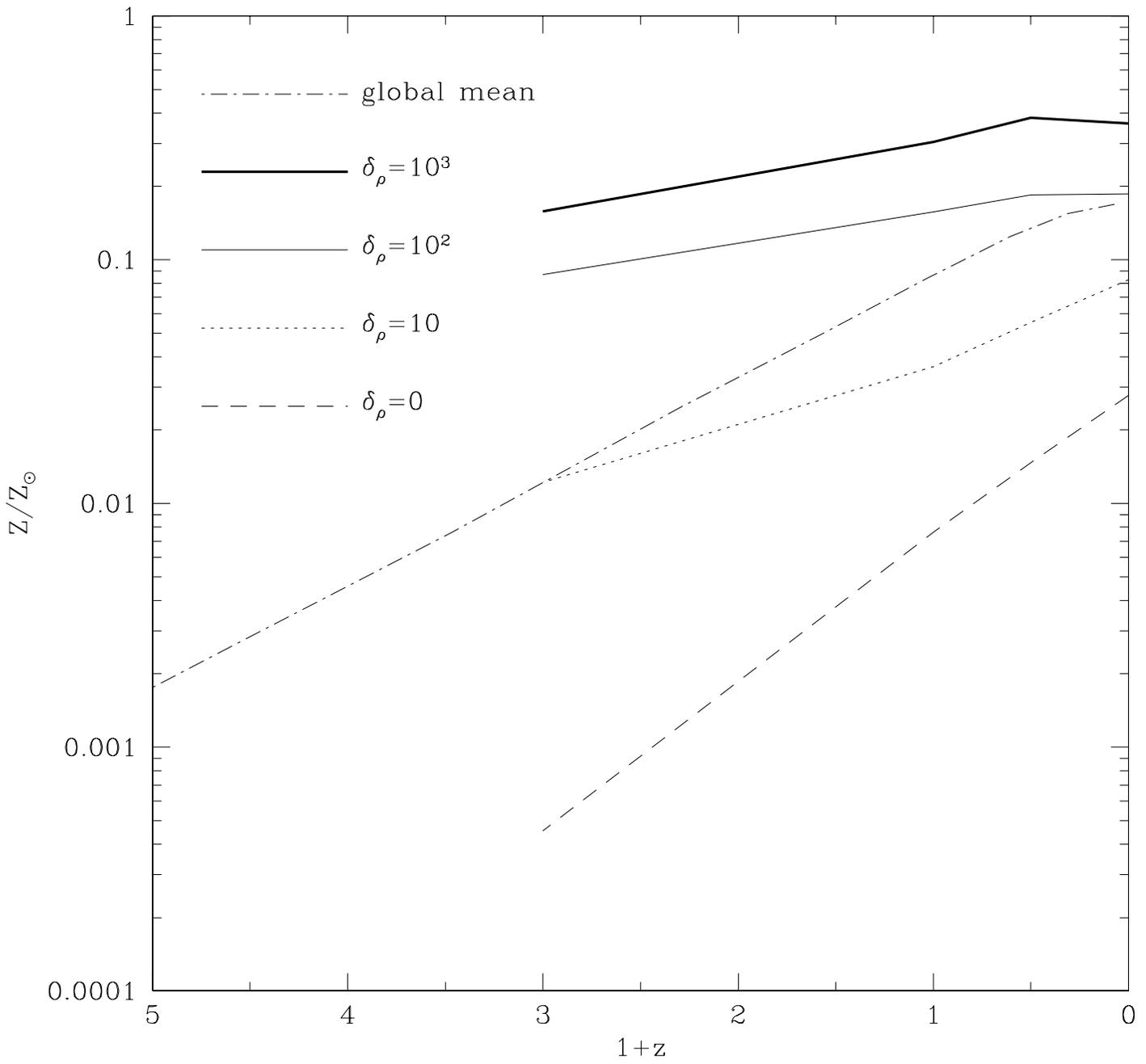
\noindent Fig. 1.--- The average metallicities averaged
over the whole universe (dot-dashed curve),
overdensity $10^3$ (thick solid curve),
overdensity $10^2$ (thin solid curve),
overdensity $10$ (dotted curve) and
overdensity $0$ (dashed curve), respectively,
as a function of redshift.

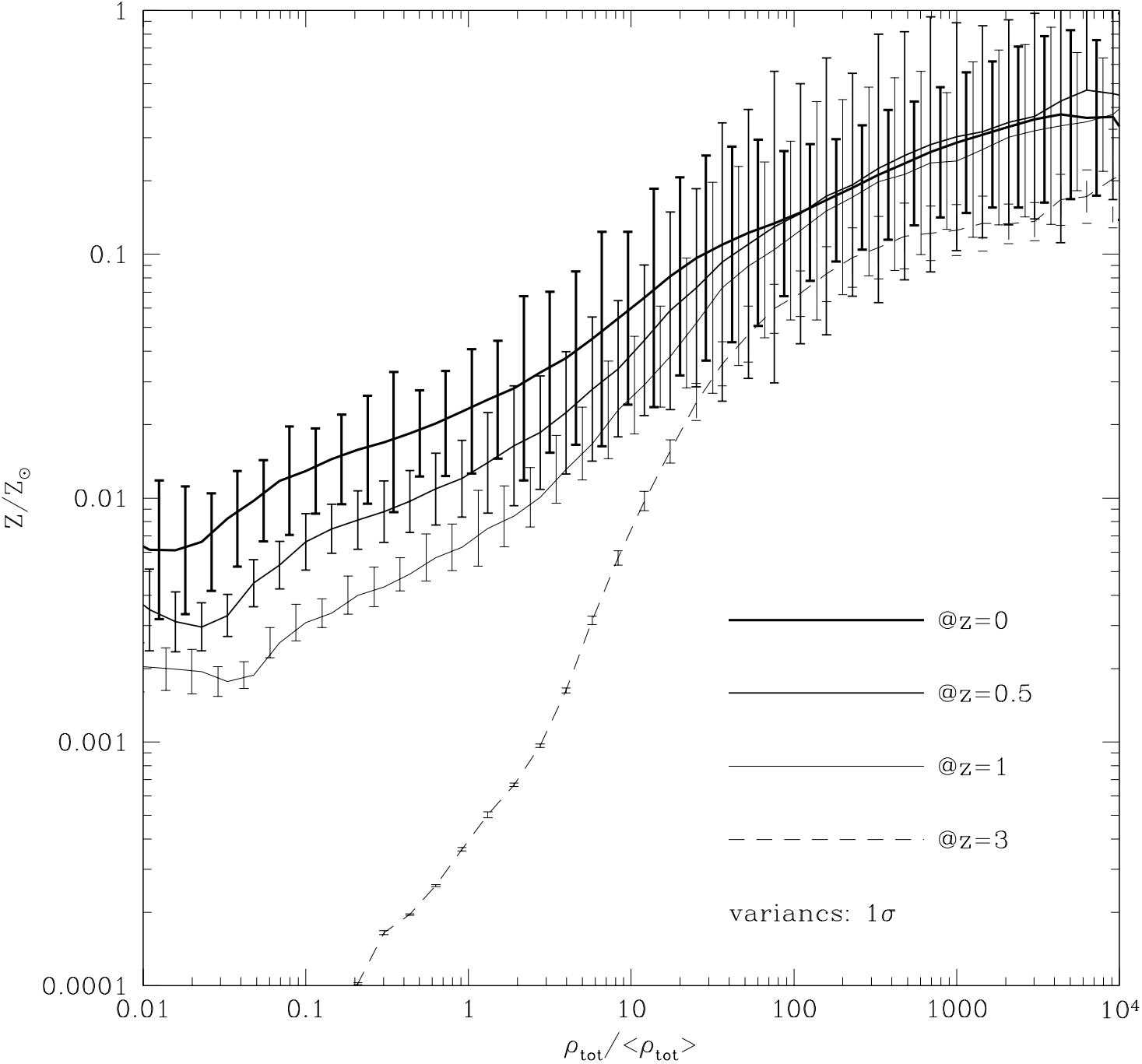
\noindent Fig. 2.--- The average metallicities as a function
of overdensity at four redshifts.
The variances are $1\sigma$.

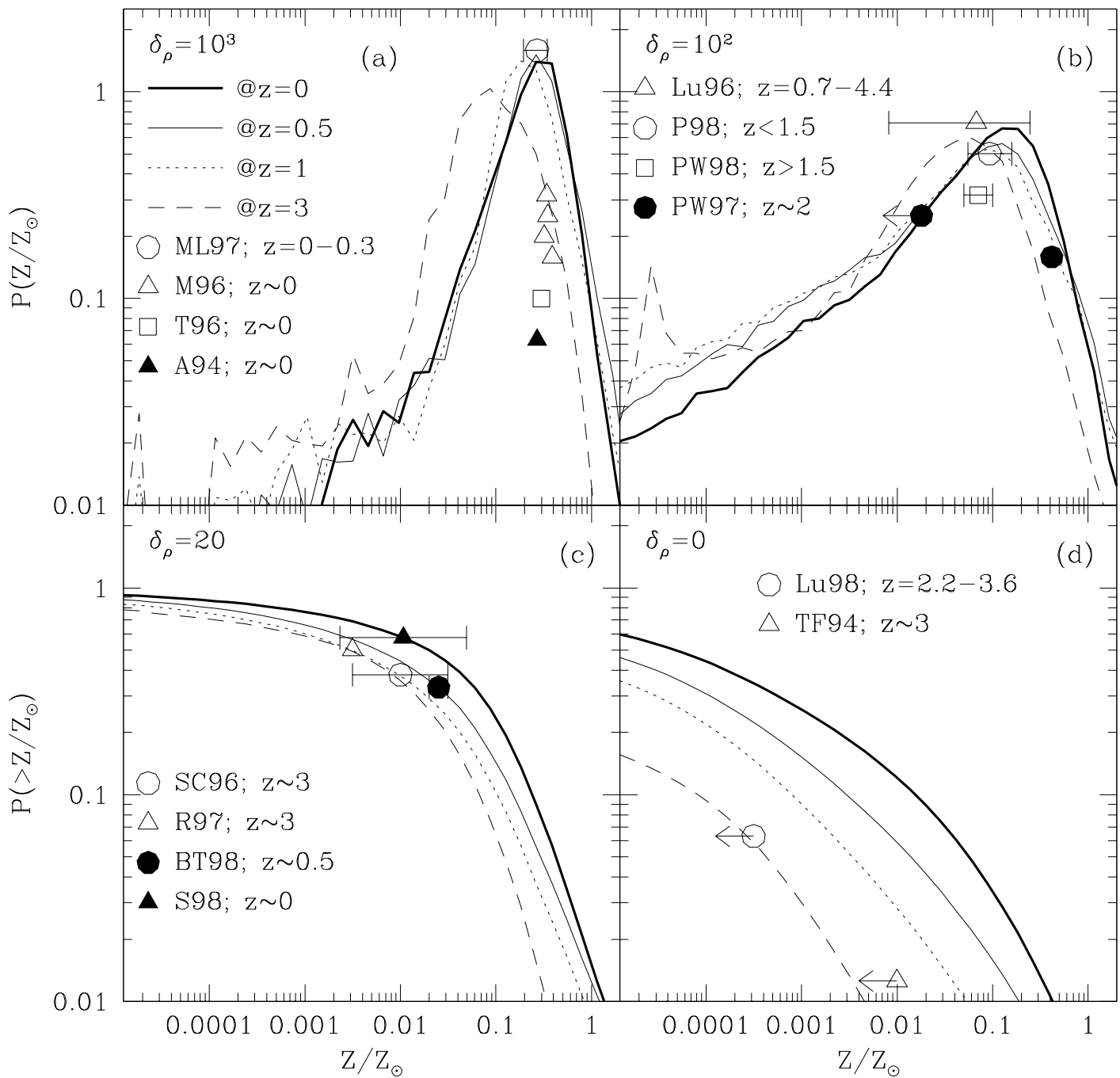
\noindent Fig. 3.---
Panel (a) 
shows the differential metallicity distribution 
for regions with overdensity $10^3$ 
(clusters of galaxies)
at four different redshifts,
$z=0$ (thick solid curve), $z=0.5$ (thin solid curve),
$z=1$ (dotted) and $z=3$ (dashed curve)
[the same convention will be used for panels (b,c,d)].
Also shown as symbols are observations from various sources.
Various symbols are
observations:
the open circle 
from Mushotzky \& Lowenstein (1997; ML97)
showing that there is almost no evolution in the intracluster
metallicity from $z=0$ to $z\sim 0.3$ at around one-third of solar,
the open triangles from from Mushotzky \etal (1996; M96)
showing the metallicities of four individual nearby clusters
(Abell 496, 1060, 2199 and AWM 7),
the open square from Tamura \etal (1996; T96) 
showing the metallicity of the intracluster gas of Abell 1060,
the filled triangle from Arnaud \etal (1994; A94) 
showing the metallicity of the intracluster gas of the Perseus cluster. 
All metallicities are measured in [Fe/H].
Panel (b) shows the differential metallicity distribution 
for regions with overdensity $10^2$.
The open triangle from Lu \etal (1996; Lu96)
shows the result from an extensive analysis of a large database of damped 
Lyman alpha systems with $0.7<z<4.4$.
The horizontal range on the open triangle
does not indicate the errorbar on the 
mean rather it shows the range of metallicities of the observed
damped Lyman alpha systems as given by Lu96.
The open circle from Pettini \etal (1998; P98)
is due to an analysis of ten damped Lyman alpha systems at $z<1.5$;
here the horizontal range indicates the error on the mean.
The open square due to Prochaska \& Wolfe (1998; PW98)
is from an analysis of 19 damped Lyman alpha systems at $z>1.5$;
the horizontal range indicates the error on the mean.
Finally, the two 
solid dots are from Prochaska \& Wolfe (1997; PW97)
of an analysis of two damped Lyman alpha systems at $z\sim 2.0$
with one having extreme low metallicity and the other having
extreme high metallicity. 
All metallicities are measured in [Zn/H].  
Vertical position in panel (b)
is without significance.
Panel (c) shows the cumulative metallicity distribution 
for regions with overdensity $20$.
The symbols are observations:
the open circle from SC96$^6$ 
for Lyman alpha clouds at $z\sim 3$ with column density
of $N>3\times 10^{14}$cm$^{-2}$,
the open triangle from Rauch \etal (1997; R97)
for Lyman alpha clouds at $z\sim 3$ with column density
of $N>3\times 10^{14}$cm$^{-2}$,
the solid dot from Barlow \& Tytler (1998; BT98)
for Lyman alpha clouds at $z\sim 0.5$ with column density
of $N>3\times 10^{14}$cm$^{-2}$,
the solid triangle from Shull \etal (1998; S98)
for Lyman alpha clouds at $z\sim 0$ with column density
of $N=(3-10)\times 10^{14}$cm$^{-2}$.
Panel (d) shows the cumulative metallicity distribution 
for regions with overdensity $0$ (i.e., mean density).
The open circle is the upper limit
for Lyman clouds with column density $N=10^{13.5}-10^{14.0}$cm$^{-2}$
at redshift $z=2.2-3.6$
from Lu \etal (1998; Lu98).
The open triangle is the upper limit
for Lyman clouds with column density $N=10^{13.0}-10^{14.0}$cm$^{-2}$
at redshift $z\sim 3$
from Tytler \& Fan (1994; TF94).
The model seems consistent with observations of low column density
Lyman alpha clouds at high redshift.

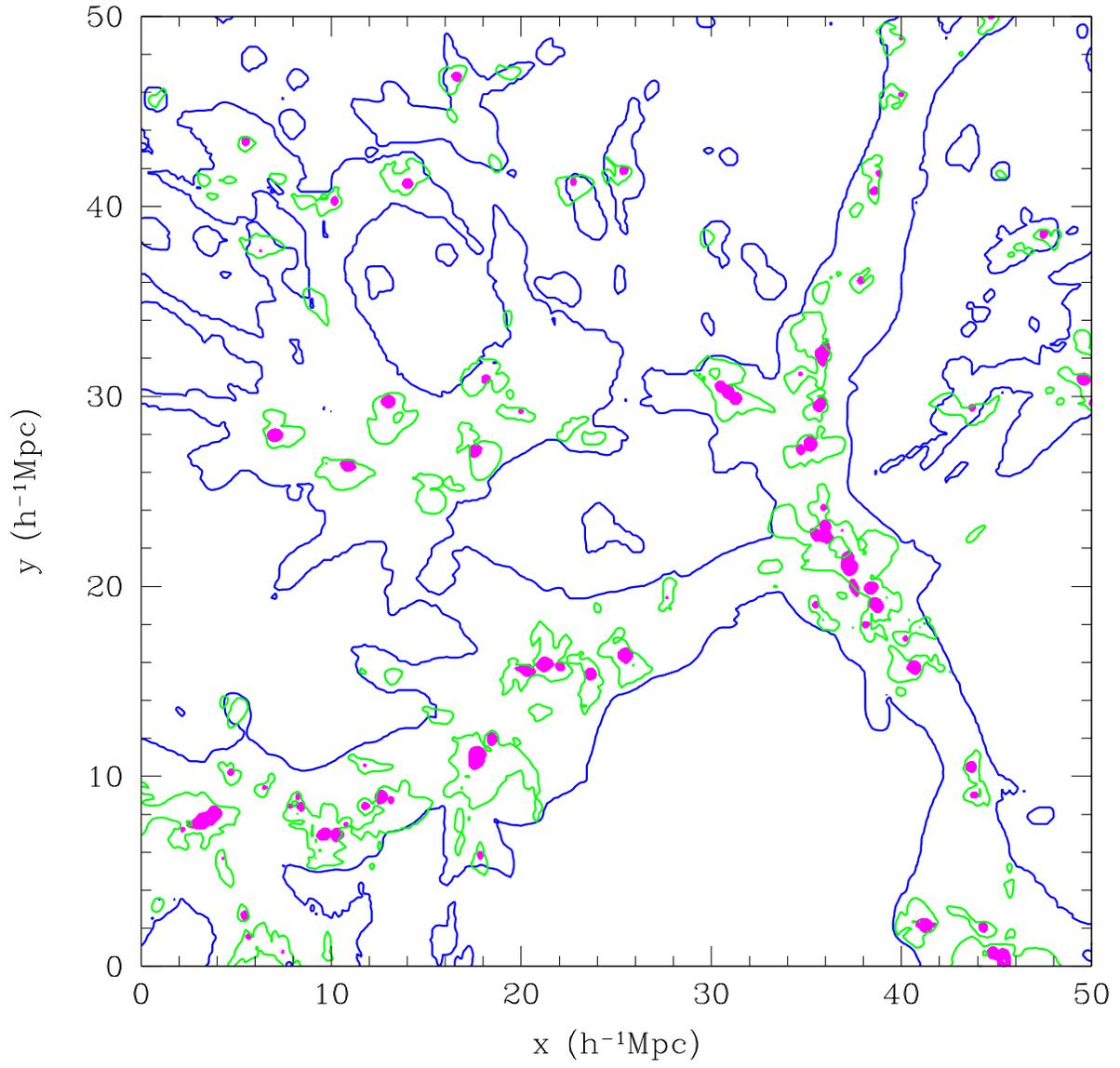
\noindent Fig. 4.---
Surface density contours on a slice
of $50\times 50\times 10h^{-3}$Mpc$^3$
for galaxies (filled red; 
at a surface density of $31$ times 
the mean surface density of galaxies),
metals (green; at a metallicity of $0.16\zsun$)
and warm/hot gas$^{12}$ with $T=10^5-10^7$K (blue;
at a surface density of $6.8$ times 
the mean surface density of warm/hot gas).
Each respective contour contains 90\% of 
the total mass of the respective component.

\end{document}